\documentclass[11pt]{article}
\usepackage{amsmath,amssymb,amsfonts,bbm}
\usepackage{multirow}
\usepackage[table,xcdraw]{xcolor}
\usepackage{graphicx}
\usepackage{caption}
\usepackage{soul} 
\usepackage{xcolor} 
\sethlcolor{yellow} 
\usepackage[numbers,sort&compress]{natbib}

\usepackage{subcaption}
\usepackage[export]{adjustbox}
\usepackage{setspace}
\usepackage{float}
\usepackage{xcolor} 
\DeclareGraphicsExtensions{.pdf,.png,.jpg}
\DeclareCaptionFormat{subfig}{\figurename~#1#2#3}
\DeclareCaptionSubType*{figure}
\usepackage{tabularx}
\newcommand{\edf}{\ {\mathop{=}\limits^{\rm def.}}\ }
\usepackage{xcolor}
\definecolor{darkyellow}{rgb}{0.72, 0.53, 0.04}

\usepackage{soul} 
\sethlcolor{yellow} 
\usepackage{soul}
\usepackage{xcolor}
\definecolor{myyellow}{RGB}{255, 250, 120}  
\sethlcolor{myyellow}

\usepackage{hyperref}
\usepackage{setspace}
\newcommand{\ee}{\end{equation}}
\newcommand{\bea}{\begin{eqnarray}}
\newcommand{\eea}{\end{eqnarray}}
\newcommand{\ba}{\begin{array}}
\newcommand{\ea}{\end{array}}

\long\def\symbolfootnote[#1]#2{\begingroup%
\def\thefootnote{\fnsymbol{footnote}}\footnote[#1]{#2}\endgroup}
\newcommand{\h}[1]{\mathop{\lambda}\limits_{#1}\ \!\!\!}

\usepackage{hyperref}

 \vspace{1cm}
 \usepackage{url}

\begin{document}

\thispagestyle{empty}
\vspace*{60pt}  
\begin{center}
    \textbf{\Large Parameterized Evolution of the Kretschmann scalar in Friedmann–Lemaître–Robertson–Walker Cosmology with Torsion Contributions and Big Rip Model}
\end{center}

\vspace{40pt}

\begin{center}
    \textbf{Tahia F. Dabash$^{3, 2}$\symbolfootnote[2]{\tt tahia.dabash@science.tanta.edu.eg}, M.A. Bakry$^{1, 2}$\symbolfootnote[1]{\tt mohamedbakry928@yahoo.com}}
\end{center}

\vspace{3pt}

\begin{flushleft}
    \begin{spacing}{1.0}
        \hspace{0.5in} $^1$ \textit{\small Mathematics Department, Faculty of Education, Ain Shams University, Cairo11864, Egypt}\\
        \hspace{0.5in} $^2$ \textit{\small Egyptian Relativity Group (ERG), Cairo University, Giza 12613, Egypt }\\
        \hspace{0.5in} $^3$ \textit{\small Mathematics Department, Faculty of Science, Tanta University, Egypt}
    \end{spacing}
\end{flushleft}

\vspace{4pt}

\begin{abstract}
We investigate the Kretschmann scalar within the framework of Parameterized Absolute Parallelism geometry, extending the classical understanding of spacetime curvature in General Relativity by incorporating torsion. This extension introduces a dimensionless parameter \( b \), allowing a seamless interpolation between RG (\( b = 0 \)) and Weitzenböck geometry (\( b = 1 \)). Using the pseudo-Riemannian metric associated with Friedmann--Lemaître--Robertson--Walker cosmology, we derive an explicit expression for the modified Kretschmann scalar, which captures contributions from standard curvature, curvature--torsion interactions, and pure torsion. Our analysis reveals that the modified scalar reduces to the classical value under specific conditions (\( b = 0 \) or \( b = \pm \sqrt{2} \)), while deviations occur for other \( b \) values. This work highlights the potential role of torsion in early universe dynamics and provides a framework for exploring deviations from standard General Relativity in cosmological models, such as the Big Bang--Big Rip model (BRM).
\end{abstract}

\section{Introduction}

The Kretschmann scalar (KS) is a fundamental geometric invariant in General Relativity (GR), providing a measure of spacetime curvature and serving as a diagnostic tool for identifying singularities and studying the formation of large-scale structures in the universe. In the standard framework of GR, which is based on purely Riemannian Geometry (RG), the concept of torsion is absent. However, torsion, a geometric property where the connection is not necessarily symmetric, arises naturally in more generalized geometric frameworks, such as those encountered in theories of gravity beyond GR. Early attempts to incorporate torsion into the geometric structure of spacetime have been made within the context of non-Riemannian geometries, notably through the study of the Weitzenböck geometry, which uses a connection with torsion. These efforts aim to address the limitations of RG in describing phenomena where torsion might play a role, such as in spinor fields and quantum gravity \cite{Cai2023,Maluf2019,Bahamonde2021,Krssak2019,Golovnev2018,Aldrovandi2020,Capozziello2021}.

To address this limitation, we adopt the Prametrized Absolute Parallelism (PAP) approach, which extends the geometric framework of GR by incorporating torsion through a dimensionless parameter \( b \). This parameter enables a smooth transition between the classical RG (\( b = 0 \)) and Weitzenböck geometry (\( b = 1 \)), allowing simultaneous contributions from curvature and torsion. By introducing this interpolation, PAP geometry provides a unified framework to explore deviations from GR and investigate the role of torsion in cosmological settings. 
The concept of interpolating between Riemannian and Weitzenböck geometries using a parameter $b$ presents an intriguing avenue for exploration in the field of theoretical physics, particularly in cosmology. Riemannian geometry, characterized by its metric tensor, describes the curvature of spacetime without torsion, while Weitzenböck geometry incorporates torsion, providing a different perspective on gravitational interactions. The novelty of this approach lies in its potential to bridge the two geometrical frameworks, allowing for a more comprehensive understanding of the role of torsion in cosmological models.

By varying the parameter $b$ , one can systematically transition between the two geometries, facilitating the examination of how torsion affects cosmic dynamics and structure formation. To articulate this novelty more clearly, it is essential to delineate how this interpolation differs from previous studies that have addressed torsion in cosmology. Prior research has primarily focused on isolated models that either adhere strictly to Riemannian or Weitzenböck frameworks. In contrast, the proposed interpolation method offers a unified framework that can capture the effects of torsion gradually, thus enabling a more nuanced analysis of its implications for phenomena such as cosmic inflation, dark energy, and the evolution of the universe.

This approach not only enhances theoretical understanding but also has the potential to provide new insights into observational data, possibly leading to predictions that could be tested against empirical findings. By clarifying these distinctions and implications, the research can establish its significance within the broader context of gravitational theories and cosmological models.
PAP is a geometric framework that extends the concept of parallelism in differential geometry, particularly in the context of Riemannian geometry. PAP provides a way to define and analyze parallelism in a more flexible manner by incorporating parameters that can adapt to various geometric settings. In Riemannian geometry, the traditional notion of parallel lines is often tied to geodesics, which are curves that represent the shortest paths between points on a curved surface. PAP allows for a broader interpretation of parallelism by considering how curves can remain ‘parallel’ under certain conditions defined by parameters.

This concept has influenced several areas of research, including:

1. \textbf{Geometric Analysis:} Studies how the properties of manifolds change under different curvature conditions, using PAP as a tool to analyze geodesics and their behaviors.

2. \textbf{General Relativity:} Investigates the implications of PAP in the context of spacetime models, where the geometry of space can be described by Riemannian metrics.

3. \textbf{Differential Equations:} Examines the solutions of certain differential equations in geometric contexts, leveraging PAP to understand the behavior of solutions along parameterized paths.

Overall, PAP enriches the study of geometry by providing new insights into the relationships between curvature, parallelism, and the intrinsic properties of spaces. Building upon the previous discussion, the PAP geometry has been instrumental in various research endeavors, as highlighted in references \cite{ref1,ref2,ref3,ref4,ref5,ref6,ref7,ref8}. This geometry provides a framework for exploring complex relationships and solving problems in diverse fields \cite{ref9,ref10,ref11}. Researchers have leveraged PAP geometry to gain insights and develop solutions in their respective domains \cite{ref12,ref13,ref14,ref15,ref16}.

In this work, we apply the PAP framework to Friedmann--Lemaître--Robertson--Walker (FLRW) cosmology, a cornerstone of modern cosmology that describes homogeneous and isotropic universes. Using a pseudo-Riemannian metric, we derive an explicit expression for the KS, highlighting the interplay between standard curvature, torsion--curvature interactions, and pure torsion contributions. This analysis provides new insights into the dynamics of the early universe and offers a platform for exploring torsion's influence in cosmological models beyond the classical GR paradigm.

The objective of this paper is to extend the concept of the KS by incorporating the effects of torsion alongside the curvature of space. Additionally, we aim to investigate the evolution of curvature within the context of the Big Bang--BRM model of the universe. To achieve these goals, the paper is structured as follows: In Section 2, we review the foundational concepts and the geometry of Absolute Parallelism (AP), with a detailed discussion of the parameterized Weitzenböck connection in Subsection 2.1. Section 3 presents the modified KS, while Section 4 explores its application to FLRW cosmology. Section 5 compares the classical and modified KSs within the Big Bang--Big Rip framework. In Section 6, we examine the evolution of curvature in FLRW cosmology. Finally, Section 7 provides a comprehensive discussion of the key findings.

\section{Brief about PAP geometry}
The foundational elements of \textit{Building Blocks} (BB) and AP geometry are essential concepts in the study of geometric structures. BB refers to the fundamental components that form the basis for constructing geometric frameworks, while AP geometry focuses on the properties of spaces where parallel lines remain equidistant, regardless of curvature. Together, these concepts provide a comprehensive understanding of the geometric landscape, enabling researchers to explore various mathematical and physical phenomena.

The study of BB and AP geometry is crucial for advancing theories in differential geometry, general relativity, and other fields that rely on the intricate relationships between curvature and parallelism in spacetime \cite{2000}.

Tetrad vectors, also known as \textit{vierbein} or \textit{frame fields}, play a crucial role in AP geometry. In this geometric framework, tetrad vectors provide a way to connect the abstract mathematical formulation of spacetime with physical concepts, particularly in the context of general relativity and theories involving torsion. Tetrad vectors consist of a set of four linearly independent vectors at each point in spacetime. They serve as a local basis for the tangent space, allowing for the representation of physical quantities in a way that is independent of the choice of coordinate system.

At each point, we define four contravariant vectors $\h{i}^{\mu}$, where ($\mu = 0,1,2,3$ represents the coordinate components and $i = 0,1,2,3$ denotes the vector index). From these, we can derive four covariant vectors $\h{i}_{\alpha}$, defined as the normalized cofactor of $\h{i}^{\mu}$ with respect to its norm. Additionally, one can establish a set of four covariant vectors that are conjugate to $\h{i}^{\alpha}$, as indicated in reference \cite{1965}.

\begin{equation}
\h{i}^{\mu}\h{i}_{\nu}\edf
\delta^{\mu}_{\nu},
\end{equation} and
\begin{equation}
\h{i}^{\nu}\h{j}_{\nu}\edf\delta_{ij},
\end{equation}
In this context, $\delta^{\mu}_{\nu}$ represents the Kronecker delta, and it is important to note that Latin indices (vector indices) are always written in lowercase, neither covariant nor contravariant. The summation convention applies to repeated indices, regardless of their positions. Utilizing the vectors defined above, we can construct the fundamental tensor in RG.

In AP geometry, the tetrad vectors facilitate the study of curvature and torsion. By expressing the metric tensor in terms of tetrads, one can explore how these geometric properties affect the dynamics of fields and particles in spacetime.
\begin{equation}\label{S}g_{\mu\nu}\edf
\h{i}_{\mu}\h{i}_{\nu},\end{equation} The metric tensor is given by \cite{Wanas2002,Wanas1989}
\begin{equation}g^{\mu\nu}\edf \h{i}^{\mu}\h{i}^{\nu}.\end{equation}
The Weitzenbock linear connection is defined as \cite{2000}
\begin{equation}\label{A}
\Gamma^{\alpha}~_{\mu\nu}\edf\h{i}^{\alpha}\h{i}_{\mu,\nu}.\end{equation}

The Weitzenböck connection is a specific type of connection used in differential geometry, particularly in the context of teleparallel gravity and certain geometric theories. Unlike the Levi-Civita connection, which relies on the metric of a manifold, the Weitzenböck connection is defined using a flat connection and parallel transport. In contrast to the Levi-Civita connection, the Weitzenböck connection includes torsion. This torsional feature is crucial in theories where torsion influences the dynamics of spacetime, particularly in specific formulations of gravity.

The Weitzenböck connection can be expressed as follows~\cite{Wanas2002}:
\begin{equation} \label{F}
\Gamma^{\alpha}_{\mu\nu} = \left\{^{\alpha}_{\mu\nu}\right\} + \gamma^{\alpha}{}_{\mu\nu},
\end{equation}
where $\gamma^{\alpha}{}_{\mu\nu}$ is the contortion tensor, defined by
\begin{equation} \label{C}
\gamma^{\alpha}{}_{\mu\nu} \stackrel{\text{def}}{=} \lambda^{\alpha}_{i} \lambda^{i}_{\mu;\nu}.
\end{equation}
The torsion tensor of the Weitzenböck connection is a fundamental object in differential geometry that describes the asymmetry of the connection. Unlike the Levi-Civita connection, which is torsion-free, the Weitzenböck connection incorporates torsion, making it particularly relevant in theories that emphasize torsional effects, such as teleparallel gravity. 

The torsion tensor $\Lambda^{\lambda}_{\mu\nu}$ is defined as the difference between the Weitzenböck connection components. It quantifies the failure of parallel transport to be path-independent in a torsional setting.
 A third order-skew tensor can be defined by
\begin{equation}\label{D}
\Lambda^{\alpha}{}_{\mu\nu}\edf\gamma^{\alpha}{}_{\mu\nu}-\gamma
^{\alpha}{}_{\nu\mu}\edf\Gamma^{\alpha}{}_{\mu\nu}-\Gamma^{\alpha}{}_{\nu\mu},\end{equation}
which is the torsion tensor of the Weitzenbock connection. 
\subsection{The parameterized Weitzenbock connection}
The parameterized Weitzenböck connection is a generalization of the standard Weitzenböck connection, characterized by introducing additional parameters that allow for more flexible configurations in the geometric structure of spacetime. This connection is particularly useful in the context of teleparallel theories of gravity, where torsion plays a significant role. This connection can be defined by  \cite{2000}
\begin{equation}\label{E}
\nabla^{\alpha}{}_{\mu\nu}=\{^{\alpha}_{\mu\nu}\}+b\gamma^{\alpha}{}_{\mu\nu},\end{equation}
where $b$ is a dimensionless parameter\cite{Wanas2002}.
The incorporation of torsion into the affine connection Eq.~\eqref{E}leads to the non-commutativity of the covariant derivatives of a covariant vector $A_{\alpha}$.
 Nevertheless, the curvature tensor continues to be defined in the conventional manner, utilizing the full connection instead of the Levi-Civita connection \cite{Wanas2007}.

\begin{equation}\label{7}
B^{\alpha}_{\ \mu\nu\sigma}  = \nabla^{\alpha}_{\ \mu\sigma||\nu} - \nabla^{\alpha}_{\ \mu\nu||\sigma} + \nabla^{\epsilon}_{\mu\sigma} \nabla^{\alpha}_{\epsilon\ \nu} - \nabla^{\epsilon}_{\mu\nu} \nabla^{\alpha}_{\epsilon \sigma}. 
\end{equation}

It can be shown \cite{Wanas2002} that the curvature Eq.\eqref{7} is related to the Riemann-Christoffel curvature $R^{\alpha}_{\ \mu\nu\sigma}$ by the relation
\begin{equation}
B^{\alpha}_{\ \mu\nu\sigma}  = R^{\alpha}_{\ \mu\nu\sigma} + b  Q^{\alpha}_{\ \mu\nu\sigma}, 
\end{equation}

and 
\begin{equation}
Q^{\alpha}_{\mu\nu\sigma} = \gamma^{\alpha_{+}}_{\mu_{+} \nu_{+}||\sigma} - \gamma^{\alpha_{+}}_{\mu_{+} \sigma_{-}||\nu}  + \gamma^{\beta}_{\mu \sigma} \gamma^{\alpha}_{\beta \nu} - \gamma^{\beta}_{\mu \nu} \gamma^{\alpha}_{\beta \sigma}.
\end{equation}

 The stroke \( || \) and the \( + \) sign are used to characterize covariant derivatives using the linear connection Eq. \eqref{E}, while the \( - \) sign is used to characterize the linear connection dual to  Eq. \eqref{E}. The geometry depending on the connection Eq. \eqref{E}, with all possible values of the parameter \( b \), is called PAP geometry.

\section{Modified KS}

In the classical framework of GR, the KS \( K_{\text{classical}} \) is a measure of spacetime curvature and is defined as the square of the Riemann curvature tensor \cite{Kretschmann1917,Wald1984,Misner1973,Perlick2004,Carroll2004}

\begin{equation}\label{k}
K_{\text{classical}} = R^\alpha_{\mu\nu\sigma} R_\alpha^{\mu\nu\sigma},
\end{equation}

Where \( R^\alpha_{\mu\nu\sigma} \) is the Riemann curvature tensor, which is derived from the metric tensor and its derivatives. This scalar is used to identify singularities in spacetime, particularly in regions of high curvature, such as near black holes or the Big Bang.

The modified KS \( K \) in this framework, incorporating torsion, is expressed using the parameterized curvature~\eqref{7}.

\begin{equation}
K = B^\alpha_{\mu\nu\sigma} B_\alpha^{\mu\nu\sigma},
\end{equation}

where \( B^\alpha_{\mu\nu\sigma} \) is the parametrized curvature tensor. Substituting the parameterized curvature tensor:

\begin{equation}
B^\alpha_{\mu\nu\sigma} = R^\alpha_{\mu\nu\sigma} + b Q^\alpha_{\mu\nu\sigma},
\end{equation}

we have

\begin{equation}
K = \left( R^\alpha_{\mu\nu\sigma} + b Q^\alpha_{\mu\nu\sigma} \right) \left( R_\alpha^{\mu\nu\sigma} + b Q_\alpha^{\mu\nu\sigma} \right).
\end{equation}

Expanding the product

\begin{equation}
K = R^\alpha_{\mu\nu\sigma} R_\alpha^{\mu\nu\sigma} + b R^\alpha_{\mu\nu\sigma} Q_\alpha^{\mu\nu\sigma} + b Q^\alpha_{\mu\nu\sigma} R_\alpha^{\mu\nu\sigma} + b^2 Q^\alpha_{\mu\nu\sigma} Q_\alpha^{\mu\nu\sigma}.
\end{equation}

Now, we can combine the two middle terms, \( b R^\alpha_{\mu\nu\sigma} Q_\alpha^{\mu\nu\sigma} \) and \( b Q^\alpha_{\mu\nu\sigma} R_\alpha^{\mu\nu\sigma} \), because they are the same. The indices are contracted in the same way, and since \( R^\alpha_{\mu\nu\sigma} Q_\alpha^{\mu\nu\sigma} \) is symmetric with respect to the exchange of \( R \) and \( Q \), the terms are identical

\begin{equation}
b R^\alpha_{\mu\nu\sigma} Q_\alpha^{\mu\nu\sigma} + b Q^\alpha_{\mu\nu\sigma} R_\alpha^{\mu\nu\sigma} = 2b R^\alpha_{\mu\nu\sigma} Q_\alpha^{\mu\nu\sigma}.
\end{equation}

Thus, we have

\begin{equation}\label{0}
K = R^\alpha_{\mu\nu\sigma} R_\alpha^{\mu\nu\sigma} + 2b R^\alpha_{\mu\nu\sigma} Q_\alpha^{\mu\nu\sigma} + b^2 Q^\alpha_{\mu\nu\sigma} Q_\alpha^{\mu\nu\sigma}.
\end{equation}
The equation \eqref{0} represents the generalized form of the KS \eqref{k}. It is noteworthy that if we set \( b = 0 \) in Eq. \eqref{0}, the KS reduces to its classical form.

The three terms represent the contributions from:

\begin{itemize}
    \item \( R^\alpha_{\mu\nu\sigma} R_\alpha^{\mu\nu\sigma} \): Standard curvature (the classical KS),
    \item \( 2b R^\alpha_{\mu\nu\sigma} Q_\alpha^{\mu\nu\sigma} \): Interaction between curvature and torsion,
    \item \( b^2 Q^\alpha_{\mu\nu\sigma} Q_\alpha^{\mu\nu\sigma} \): Pure torsion.
\end{itemize}

\section{Application to FLRW Cosmology}
In order to study the physical initial singularities of homogeneous and isotropic world models, we use the modified KS as a tool for analysis. To this end, we need to work within certain PAP structures that satisfy the necessary symmetry conditions, namely homogeneity and isotropy. When studying such singularities in the context of standard cosmology in GR, the FRW-Riemannian structure is commonly employed.
The selection of tetrads is a fundamental aspect of formulating theories in geometrical frameworks, particularly in the context of the FLRW metric. Tetrads, or \textit{vierbeins}, serve as a bridge between the curved spacetime described by general relativity and the local flat Minkowski space. They enable the description of the gravitational field in a way that incorporates torsion and other geometric features.

In relation to the FLRW metric, which models a homogeneous and isotropic universe, the choice of tetrads can significantly affect the interpretation of physical quantities such as energy density and pressure. By appropriately selecting the tetrads, one can derive the effective metric that governs the dynamics of cosmological models, allowing for a comprehensive analysis of how different geometrical properties influence cosmic evolution.

Furthermore, the tetrad formalism provides a powerful tool for exploring the implications of torsion and other modifications to standard gravity. It allows for a more flexible representation of the underlying geometry, making it possible to investigate the effects of torsion on cosmological dynamics and structure formation. Thus, understanding the choice of tetrads and their relationship to the FLRW metric not only enriches theoretical insights but also has profound implications for observational cosmology.
To ensure consistency with the FLRW metric, we select a diagonal tetrad field that preserves both homogeneity and isotropy.This tetrad choice simplifies the geometry and guarantees that the derived connection and curvature scalars reflect the underlying symmetries of the spacetime. It also allows for the torsion parameter $ b $ to be incorporated naturally via the PAP formalism.

This structure conforms to the cosmological principle, exhibiting both homogeneity and isotropy. In the current study, we employ either an AP or PAP structure, both of which also adhere to the cosmological principle. Since the AP and PAP structures share identical BB, we utilize the BB from the AP structure that aligns with the cosmological principle as articulated by Robertson in 1932~\cite{Robertson1932}. 
\begin{equation}\label{v}
\lambda_i^\mu = 
\begin{pmatrix}
1 & 0 & 0 & 0 \\
0 & L^+ \sin \theta \cos \phi & \frac{4}{4A} \sin \theta \sin \phi & L^- \cos \theta \cos \phi - 4 \sqrt{k r} \cos \theta \\
0 & L^+ \sin \theta \sin \phi & \frac{4}{4A} \cos \theta & L^- \cos \theta \sin \phi + 4 \sqrt{k r} \sin \theta \\
0 & L^+ \cos \theta & \frac{4}{4A} & - L^- \sin \theta
\end{pmatrix}
\end{equation}
where \( L^+ = 4 + k r^2 \), \( L^- = 4 - k r^2 \), \( k \) is the sectional curvature constant taking the values (+1, 0, -1), and \( A \) is an unknown function of \( t \) only. The associated  pseudo-Riemannian metric to the matrix \eqref{v} is
\begin{equation}
ds^2 = dt^2 - \frac{16A^2}{L^{+2}} \left( dr^2 + r^2 d\theta^2 + r^2 \sin^2\theta d\phi^2 \right),
\end{equation}

The non-zero components of the parameterized curvature tensor Eq. \eqref{7} have been computed previously \cite{Wanas2018}. Therefore, we will use these components to derive the modified KS in the PAP- geometry.

The non-zero components of \( 
B_{\mu\nu} = B^\alpha_{\ \mu\alpha\nu} = g^{\alpha\beta} B_{\alpha\mu\beta\nu}.
 \) 
 are \cite{Wanas2018}
\begin{eqnarray}
B_{00} &=& 3(b - 1)\frac{\ddot{A}}{A}, \\
B_{11} &=& -16(b - 1)\frac{2(1 - b)\dot{A}^2 + 2k(1 + b) + A\ddot{A}}{L^2}, \\
B_{22} &=& -16r^2(b - 1)\frac{2(1 - b)\dot{A}^2 + 2k(1 + b) + A\ddot{A}}{L^2}, \\
B_{33} &=& -16r^2\sin^2\theta(b - 1)\frac{2(1 - b)\dot{A}^2 + 2k(1 + b) + A\ddot{A}}{L^2}.
\end{eqnarray}
where \( \dot{A}=\frac{dA}{dt} \) .
We substitute these components into the expression:
\begin{equation}
 K = g^{00} g^{00} B_{00}^2 + g^{11} g^{11} B_{11}^2 + g^{22} g^{22} B_{22}^2 + g^{33} g^{33} B_{33}^2,   
\end{equation}

where the inverse metric components are:
\begin{equation}
   g^{00} = 1, \quad g^{11} = -\frac{L^2}{16A^2}, \quad g^{22} = -\frac{L^2}{16A^2 r^2}, \quad g^{33} = -\frac{L^2}{16A^2 r^2 \sin^2\theta}. 
\end{equation}

Substituting the inverse metric components and the curvature tensor components, the KS becomes:
\begin{equation}
  K = B_{00}^2 + \frac{L^4}{(16A^2)^2} B_{11}^2 + \frac{L^4}{(16A^2 r^2)^2} B_{22}^2 + \frac{L^4}{(16A^2 r^2 \sin^2\theta)^2} B_{33}^2.  
\end{equation}

Finally, substituting the expressions for \( B_{00}, B_{11}, B_{22}, \) and \( B_{33} \), we obtain the explicit form of the KS for FLRW cosmology as a function of \( A(t), \dot{A}(t), \ddot{A}(t), \) and the parameter \( b \):
\begin{equation}
    K = \left[ 3(b-1)\frac{\ddot{A}}{A} \right]^2 + \ldots
\end{equation}

where the full expression is derived through symbolic expansion and simplification.
he modified KS is given by

\begin{equation}\label{l}
K_{\text{modified}} = \frac{12 (b-1)^2 (1 + b)^2 k^2}{A[t]^4}.
\end{equation}

The discussion regarding how torsion, mediated by the parameter $b$, influences observable cosmological dynamics is crucial for understanding the broader implications of this theory. Torsion introduces an additional degree of freedom in the geometric description of spacetime, which can lead to significant modifications in the behavior of cosmic structures and their evolution.

A key aspect of this analysis is the modified Kretschmann scalar, which serves as a measure of curvature in a gravitational theory. In traditional Riemannian geometry, the KS is a well-defined quantity that indicates the presence of curvature due to matter and energy. However, when torsion is introduced, this scalar may be altered, reflecting the new geometrical features imparted by torsion.

Physically, the modifications to the KS can indicate changes in the gravitational field's strength and behavior, potentially affecting phenomena such as the expansion rate of the universe, the formation of cosmic structures, and the dynamics of dark matter and dark energy. A clearer insight into these modifications allows for a deeper understanding of how torsion might influence observable quantities such as the cosmic microwave background radiation, galaxy formation, and gravitational waves.
Also, the modified Kretschmann scalar  $K_{\text{modified}}$  provides a measure of how spacetime curvature responds to torsion effects in the cosmological setting. Unlike the classical scalar in general relativity, $K_{\text{modified}}$  incorporates the torsion parameter  $b$ ,  which encodes deviations from Riemannian geometry. This modified curvature affects gravitational strength and tidal forces, especially in the early or late-time universe. Its evolution may influence observable features like structure formation, cosmic acceleration, or gravitational wave dynamics. Understanding how  $K_{\text{modified}}$  changes with $b$   allows us to assess the phenomenological relevance of torsion in cosmological observations.

From Eq.~\eqref{l}, we make the following observations:
We can observe the following from \eqref{l}
\begin{itemize}
        \item At \( b = \pm \sqrt{2} \), \( K_{\text{modified}} \) again matches \( K_{\text{classical}} \), showing specific torsion values where contributions cancel out.
        \item For \( |b| > \sqrt{2} \) or \( |b| < \sqrt{2} \), \( K_{\text{modified}} \) deviates significantly, increasing rapidly with larger \( |b| \).

        \item The curve is symmetric about \( b = 0 \), as the torsion factor \( (-1 + b)^2 (1 + b)^2 \) depends on \( b^2 \).
    \end{itemize}

    \begin{itemize}
        \item For \( b \neq 0, \pm \sqrt{2} \), torsion introduces additional curvature, leading to deviations from \( K_{\text{classical}} \).
        \item The deviations grow rapidly as \( |b| \) increases.
       \item When \( b = 0 \), Eq.\eqref{l} is reduced to the classical KS in RG as in the following form \cite{Gkigkitzis2014}.

    \end{itemize}

\begin{equation}
 K_{\text{classical}} = \frac{12 k^2}{A[t]^4}.   
\end{equation}

The condition for the modified scalar to return to the classical value is:
\begin{equation}
  (-1 + b)^2 (1 + b)^2 = 1.  
\end{equation}

This gives:
\begin{equation}
   b = 0 \quad \text{or} \quad b = \pm \sqrt{2}. 
\end{equation}

\section{Comparison of Classical and Modified KS in Big Rip Model}
The Big Rip Model (BRM) is a theoretical cosmological scenario that suggests the universe could end in a dramatic and catastrophic manner~\cite{ref29, ref30}. According to this model, the expansion of the universe, driven by dark energy, accelerates to the point where it eventually tears apart galaxies, stars, planets, and even atomic structures. This scenario is predicated on the assumption that the density of dark energy increases over time, leading to an exponential expansion of the universe. The key features of the BRM can be summarized as follows\\
1. Accelerating Expansion: The BRM is based on the observation that the universe's expansion is accelerating, a phenomenon attributed to dark energy. This acceleration could become so extreme that it overwhelms all other forces in the universe~\cite{ref31, ref34}.

2. Timeline: In the BRM, the universe would reach a point where the expansion rate becomes infinite. This could happen in a finite amount of time, potentially within a few billion years, depending on the properties of dark energy.

3. Destruction of Cosmic Structures: As the expansion accelerates, it would first affect galaxy clusters, pulling them apart. Eventually, galaxies would be torn apart, followed by solar systems, and finally, the very atoms that make up matter would be ripped apart.
4. Theoretical Basis: The BRM is one of several hypotheses regarding the ultimate fate of the universe, alongside others like the Big Crunch and Heat Death. It is primarily derived from models of dark energy and its equation of state \cite{ref29}.

The scaling factor that characterizes the BRM can be expressed as follows \cite{ref35, ref36}:

\begin{equation}
  A(t) = \left(\frac{t}{2m - a t}\right)^{1/m},   
\end{equation}

where \( A_0 \) is a constant of integration and can be taken as a unity, \( a \) and \( m \) are constants. 

The model of the universe starts with a Big Bang model at \( t_{bb} = 0 \) and ends with a BRM at \( t_{br} = \frac{2m}{a} \) \cite{ref35}.

By substituting from Eq. (46) into Eq. (41), we obtain:

\begin{equation}
 K_{mod.}(t) = 12(b - 1)^2 (b + 1)^2 k^2 \left(\frac{2m - a t}{t}\right)^{1/m}.   
\end{equation}
 
\section{Curvature Evolution in FLRW Cosmology}

In the context of FLRW cosmology, the evolution of curvature with torsion is significant for understanding the dynamics of the universe, particularly in models like the BRM. The KS, which quantifies the curvature of spacetime, plays a crucial role in analyzing singularities and the behavior of gravitational fields under the influence of torsion. In the BRM scenario, the rapid expansion of the universe leads to flatness of the universe, potentially resulting in the breakdown of physical laws as we approach the singularity. This interplay of curvature and torsion provides deeper insights into the nature of cosmic evolution and the fate of the universe.
In our research, we can select coefficient values to investigate the impact of gravity and torsion on the KS for the BRM. For instance, as demonstrated in Refs.~\cite{ref35, ref36}, we might choose \( a = 0.126 \), \( m = 2 \), and \( k = \pm 1 \).
In Fig.1, we illustrate the curvature of space-time in the BRM. The plots clearly show how increasing the torsion parameter $b$ suppresses the growth of curvature over time. The parameter $b$ serves to diminish the curvature as the universe approaches the Big Rip event. This reduction becomes more pronounced for larger values of $b$, suggesting that torsion plays a smoothing role in late-time cosmological dynamics.This indicates that the curvature was initially very high at the beginning of the universe at the Big Bang, gradually decreasing until it reached a flat state right at the moment of the Big Rip.
The Kretschmann scalar (30), a crucial measure in general relativity, quantifies the degree of curvature of spacetime. It is derived from the Riemann curvature tensor and serves as an important indicator of gravitational field strength. In scenarios leading to the Big Rip—a theoretical end of the universe characterized by accelerated expansion—the values of the Kretschmann scalar can indicate extremely low gravitational conditions. One can see from Fig.~1 that, as the universe progresses, the interplay of curvature and torsion may result in decreasing values of the Kretschmann scalar, indicating areas of significant gravitational stress. Ultimately, these values could approach zero, signaling a potential breakdown of spacetime integrity as it nears the singularity linked to the Big Rip. This scenario suggests that the established laws of physics may no longer apply, leading to a disintegration of the very fabric of spacetime.

The following graph illustrates the comparison between the classical KS (\( K_{\text{classical}} \)) and the modified KS (\( K_{\text{modified}} \)) as a function of the torsion parameter \( b \).

\begin{figure}[H]
    \centering
    \begin{subfigure}[b]{0.45\textwidth}
        \includegraphics[width=\textwidth]{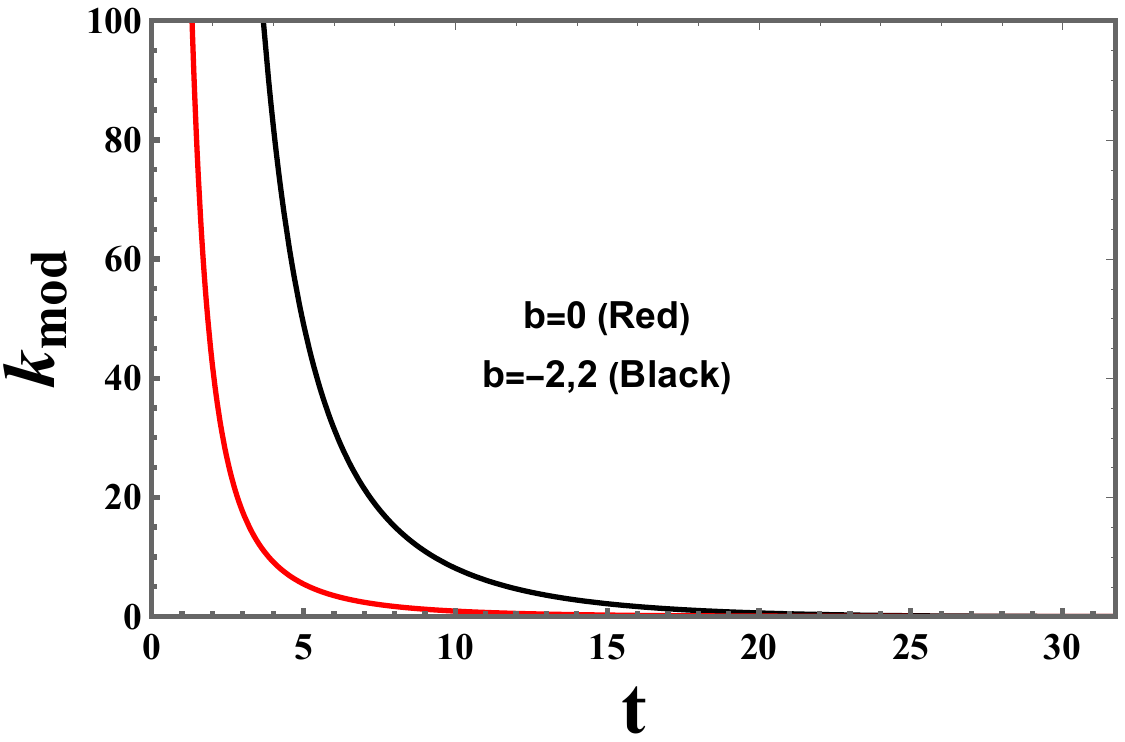}
        \caption*{\textnormal{(a)}}
    \end{subfigure}
    \hspace{0.05\textwidth}
    \begin{subfigure}[b]{0.45\textwidth}
        \includegraphics[width=\textwidth]{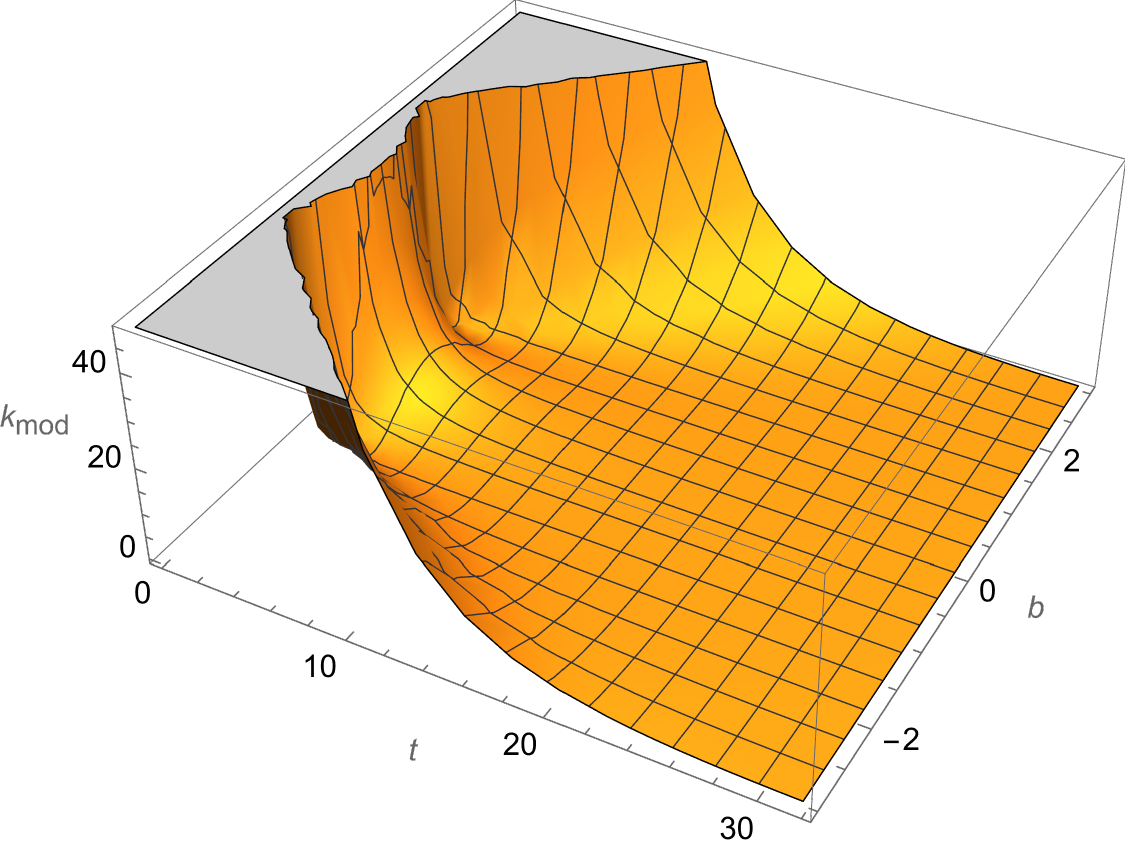}
        \caption*{\textnormal{(b)}}
    \end{subfigure}
    
    \caption{(a) Time evolution of the curvature $(K_\text{mod} $in a BRM model for different values of the torsion parameter $b$. $b$ Effect of torsion parameter $b$ on the curvature scalar $K_\text{mod}$, showing that higher values of $b$ lead to reduced curvature.}
    \label{fig:curvature_brm}
\end{figure}

\begin{figure}[H]
    \centering
    \includegraphics[width=0.7\textwidth]{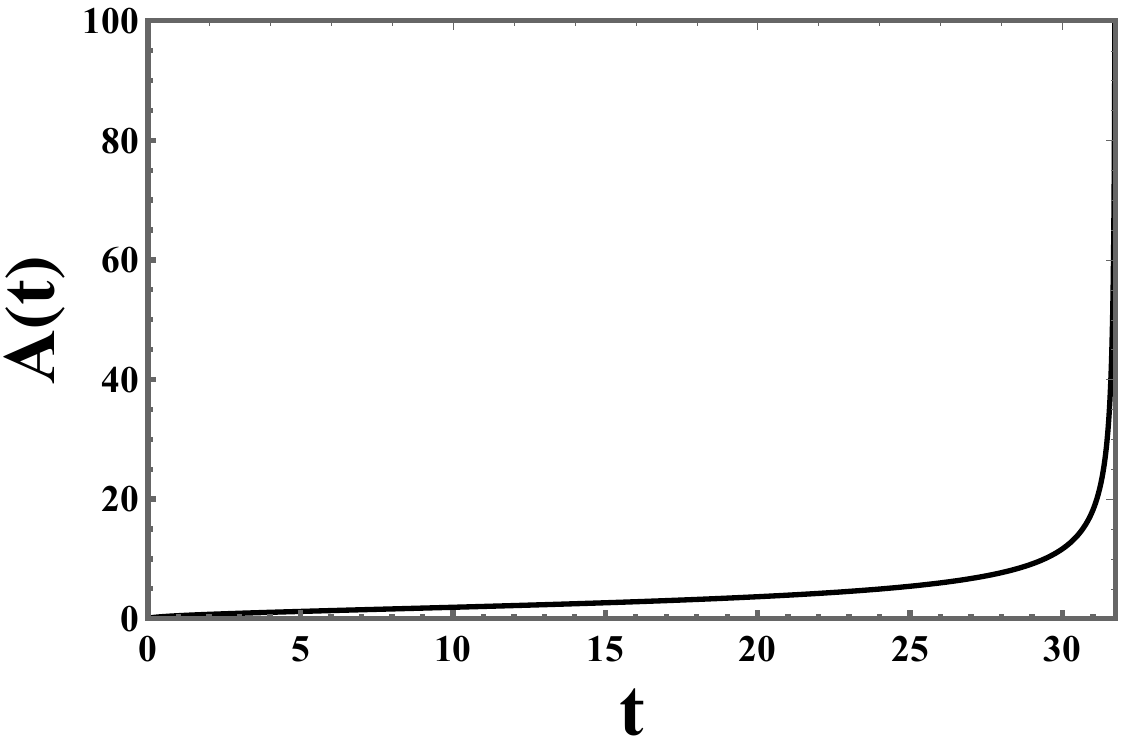}
    \caption{Time evolution of the scale factor $A(t)$ in the BRM model, diverging at the Big Rip ($t = \frac{2m}{a}$).}
    \label{fig:big_rip_scale}
\end{figure}

In Fig.~\ref{fig:big_rip_scale}, the relationship between the cosmic scale factor and time is illustrated for the Big Rip scenario. It is evident that the universe originated from a state of nothingness at the Big Bang (i.e., $A = 0$ at $t = 0$) and ultimately approaches infinity at the moment of the Big Rip (i.e., $A = \infty$ at $t = \frac{2m}{a}$).

\section{Discussion}
The study of the KS holds significant importance in the field of cosmology for several reasons. The KS is a mathematical quantity that encapsulates the curvature of spacetime. It provides insights into the geometric properties of the universe, which are crucial for understanding various cosmic phenomena. By analyzing the KS in different cosmological models, researchers can test the predictions of GR. Variations in its value can indicate the presence of singularities or anomalies within a given model, aiding in the validation or refinement of gravitational theories.
In recent studies, the incorporation of torsion and other modifications to GR has led to new insights. The modified KS can reveal how these factors influence the dynamics of the universe, potentially leading to new theories of gravity. Accordingly, in this study, we have explored the modification of the KS within the framework of PAP geometry, where torsion contributions are considered alongside standard curvature. By introducing the parameter \( b \), which interpolates between Riemannian and Weitzenböck geometries, we have derived an expression for the modified KS. The results show that the scalar reduces to the classical value under specific conditions, namely when \( b = 0 \) (no torsion) or \( b = \pm \sqrt{2} \) (where torsion contributions cancel out).
Our analysis has provided new insights into the role of torsion in cosmological models, especially in the early universe, where torsion effects could influence the dynamics of spacetime. The modified KS has been used as a diagnostic tool for exploring deviations from standard GR, and it highlights how torsion can alter the curvature structure, particularly during periods of high energy density.

The study has shown that torsion has a significant impact on spacetime curvature in the early universe, affecting the behavior of the KS and modifying the structure of singularities. While torsion diminishes in influence as the universe evolves, its presence provides a potential avenue for exploring alternative gravity theories that deviate from traditional GR. The study of the KS is essential in cosmology, particularly within the framework of the Big Bang-BRM, for several important reasons. The KS quantifies the curvature of spacetime, offering crucial insights into the geometric structure of the universe. In the BRM, where the universe undergoes accelerated expansion leading to a potential singularity, understanding this curvature becomes vital, as illustrated in Figures 1 and 2. In the Big Bang-BRM, the expansion of the universe accelerates to a point where cosmic structures ultimately disintegrate. The KS serves as a valuable tool for identifying and analyzing these singularities, enabling cosmologists to comprehend the conditions that give rise to them and their implications for the universe's fate. This analysis indicates that the curvature was initially very high at the onset of the universe during the Big Bang, gradually decreasing until reaching a flat state just before the BRM. In summary, the examination of the KS is fundamental for understanding the dynamics of the universe in the context of the BRM. It provides critical insights into spacetime curvature and singularities. Table 1 summarizes the physical analysis of the modified KS concerning the evolution of the universe within the Big Bang-Big Rip framework.
\begin{table}[H]
\centering
\caption{Physical Analysis of $K_{\text{modified}}$ Evolution}
\begin{tabular}{|m{4cm}|m{10cm}|}
\hline
\rowcolor{lightgray}
\textbf{Time Era} & \textbf{Physical Analysis} \\ \hline
\textbf{Early Times (\(t = 0\))} & 
\begin{itemize}
    \item Sharp peaks in $K_{\text{modified}}$ are observed for all values of the torsion parameter \(b\).
    \item These peaks indicate high spacetime curvature, consistent with extreme density and energy conditions near the Big Bang.
    \item For \(b = 0\) (no torsion), the curvature reaches maximum values, showing the dominance of Einsteinian gravity.
    \item For \(b > 0\) (non-zero torsion), the peaks are suppressed, suggesting that torsion reduces the singularity's intensity and moderates spacetime curvature.
\end{itemize} \\ \hline
\textbf{Late Times (\(t =\frac{2m}{a}\))} & 
\begin{itemize}
    \item The scalar \(K_{\text{modified}}\) asymptotically approaches zero, representing a flat or nearly flat spacetime as the universe becomes increasingly homogeneous and isotropic.
    \item The torsion parameter \(b\) has a diminishing impact on curvature at late times.
\end{itemize} \\ \hline
\end{tabular}

\end{table}
Future work will involve further exploration of torsion in the context of other cosmological models and its implications for the nature of spacetime at both large and small scales.

\end{document}